\documentclass[conference]{IEEEtran}
\IEEEoverridecommandlockouts

\usepackage{cite}
\usepackage{amsmath,amssymb,amsfonts}
\usepackage{algorithmic}
\usepackage{graphicx}
\usepackage{textcomp}
\usepackage{xcolor}
\usepackage{makecell}
\usepackage{multirow}
\usepackage[inline]{enumitem}
\usepackage{pifont}
\usepackage{url}
\def\BibTeX{{\rm B\kern-.05em{\sc i\kern-.025em b}\kern-.08em
		T\kern-.1667em\lower.7ex\hbox{E}\kern-.125emX}}
\begin{document}
	
	\title{Red Team Redemption: A Structured Comparison of Open-Source Tools for Adversary Emulation}
	

\author{\IEEEauthorblockN{Max Landauer, Klaus Mayer, Florian Skopik, Markus Wurzenberger, Manuel Kern}
	\IEEEauthorblockA{\textit{Center for Digital Safety \& Security} \\
		\textit{Austrian Institute of Technology}\\
		Vienna, Austria \\
		firstname.lastname@ait.ac.at}
}
	
	\maketitle
	
	\begin{abstract}
		Red teams simulate adversaries and conduct sophisticated attacks against defenders without informing them about used tactics in advance. These interactive cyber exercises are highly beneficial to assess and improve the security posture of organizations, detect vulnerabilities, and train employees. Unfortunately, they are also time-consuming and expensive, which often limits their scale or prevents them entirely. To address this situation, adversary emulation tools partially automate attacker behavior and enable fast, continuous, and repeatable security testing even when involved personnel lacks red teaming experience. Currently, a wide range of tools designed for specific use-cases and requirements exist. To obtain an overview of these solutions, we conduct a review and structured comparison of nine open-source adversary emulation tools. To this end, we assemble a questionnaire with 80 questions addressing relevant aspects, including setup, support, documentation, usability, and technical features. In addition, we conduct a user study with domain experts to investigate the importance of these aspects for distinct user roles. Based on the evaluation and user feedback, we rank the tools and find MITRE Caldera, Metasploit, and Atomic Red Team on top.
	\end{abstract}
	
	
	\section{Introduction}
	
	The ever increasing number and sophistication of cyber attacks poses many challenges to organizations \cite{mtrends2023, crowdstrike}. As a consequence, security specialists employ a wide range of countermeasures, including technical approaches for automatic monitoring and intrusion detection \cite{khraisat2019survey}, educational measures such as security awareness trainings \cite{zwilling2022cyber}, policies or strategic approaches such as incident response planning \cite{ahmad2020integration}, vulnerability management \cite{syed2020cybersecurity}, risk assessments \cite{cremer2022cyber}, and situational awareness \cite{ahmad2021can}, among many more. However, even with many diverse measures in place and functional, it is often difficult to assess their effectiveness and completeness since blind spots are easily missed \cite{skopik2022blind}.
	
	It is thus common practice to conduct regular tests that challenge security defenses of organizations. The term penetration testing generally refers to any activities related to the execution of cyber attacks for the purpose of identifying risk areas and weaknesses in applications, systems, and networks \cite{alhamed2023systematic}. The main purpose of penetration testing is to proactively recognize security issues and improve the organization's security posture before an actual attacker with malicious intent is able to intrude the network and cause any damage \cite{mughal2022building}. 
	
	Red teaming has emerged as one of the most realistic and advanced strategies to conduct comprehensive security tests. Thereby, red teams, which comprise people with diverse skills in cyber security domains and beyond, simulate adversaries and conduct planned attack exercises without informing defenders in advance \cite{kovavcevic2020red}. In contrast to vulnerability assessment and penetration testing, which only achieve to identify weaknesses and assess the risk associated with them, red teaming relies on interaction between attackers and defenders and thereby facilitates training of security teams and improvement of their detection and response capabilities \cite{vos2022capability}. 
	
	Unfortunately, the implication of such an interactive exercise is that red teaming is a mostly manual process that incurs significant costs as it involves time-consuming tasks conducted by experienced personnel \cite{miller2018automated}. To alleviate this issue, adversary emulation tools are frequently employed to automate some of the tasks in red teaming exercises, which helps practitioners to establish a baseline of security measures. Thereby, capabilities of these tools range from simple technique execution to full emulation of an adversary \cite{applebaum2017analysis, miller2018automated}. 
	
	While there is obviously some benefit in having a large pool of adversary emulation tools to choose from, it is often difficult to do so when one tries to select a suitable emulator for a specific use-case and there is only limited time to explore and experiment with several of the tools \cite{zilberman2020sok}. For example, in situations where security professionals aim to conduct attacks against very specific components and applications, suitable tools should enable the creation of custom attack procedures and not just support a predefined set of attack cases. Other use-cases could revolve around more basic attack scenarios, but require that they are executed by personnel without red teaming experience, in which case usability of the tool is more important than advanced technical capabilities. Anyway, it is necessary to consider the needs and experience of the users of adversary emulation tools to make an informed decision.
	
	There are several challenges that need to be addressed. First, it is non-trivial to identify all properties of adversary emulation tools that should be taken into account. Second, it is difficult to differentiate which properties are more relevant than others, in particular, since users in different roles may have distinct requirements and needs. Third, it is very time-consuming to review multiple tools in a hands-on manner to evaluate their appropriateness for certain use-cases. To address these challenges, we conduct a structured review and comparison of adversary emulation tools comprising (i) a literature study to collect relevant properties that we assemble into a questionnaire, (ii) technical assessment with hands-on experiments using a set of nine tools, and (iii) a user study to ascertain the importance of properties for certain user roles. In our work we thus answer the following research questions: \textit{RQ1: What properties of adversary emulation tools are the most relevant for stakeholders? RQ2: Which adversary emulation tools are best suited to fulfill the needs of certain user groups?}
	
	To the best of our knowledge, this work provides the most comprehensive review and comparison of adversary emulation tools. Other studies do not assess how important users rate certain properties of tools \cite{zilberman2020sok}, focus on detectability of tools \cite{elgh2022comparison, orbinato2024laccolith}, are limited to specific operating systems \cite{chen2020poster, stock2022faehigkeitsanalyse}, or consider fewer questions \cite{zilberman2020sok} and tools \cite{elgh2022comparison} for evaluation. Our study, on the other hand, aims at a broad comparison of open-source tools. We summarize our contributions as follows:
	
	\begin{itemize}
		\item A questionnaire capturing properties and features of adversary emulation tools,
		\item an online survey for assessment of relevance scores for various aspects of these tools, and
		\item an evaluation and ranking of publicly available tools.
	\end{itemize}
	
	The remainder of the paper is structured as follows. Section \ref{related} summarizes the background of red teaming and related works in the research field of adversary emulation tools. Section \ref{methodology} describes the methodology of our study, including tool selection, questionnaire design, survey design, tool evaluation, and scoring. Section \ref{tools} provides an overview and brief description of all selected tools. We present the results the evaluation study and online survey in Sect. \ref{evaluation} and answer our research questions in Sect. \ref{discussion}. Section \ref{conclusion} concludes the paper. We provide our questionnaire in Appendix \ref{appendix}.
	
	\section{Background \& Related Work} \label{related}
	
	The term \textit{red teaming} is often incorrectly used interchangeably with other types of security tests, in particular, penetration tests. Kova{\v{c}}evi{\'c} et al. \cite{kovavcevic2020red} thus conduct a literature study on security tests. They conclude that while penetration tests are short-term exercises that validate an organization's security posture, red teaming provides an ongoing training of security personnel. The authors also state that beside simulating real attackers, red teams also support organizations by acting as devil's advocates or consultants. 
	
	Given their definition, it is easy to understand that red teaming is a mostly manual and interactive task that is difficult to automate; nonetheless, several authors of scientific works have proposed adversary emulation tools in an attempt to take steps in that direction. For example, Plot et al. \cite{plot2020cartt} propose a Cyber Automated Red Team Tool (CARTT) that provides an easy-to-use interface to carry out vulnerability scans and obtain recommendations on how to mitigate threats based on its findings. Another example is LACCOLITH, an agent presented by Orbinato et al. \cite{orbinato2024laccolith}, that was specifically designed to evade detection by antivirus. Chen et al. \cite{chen2020poster}, on the other hand, design an adversary emulation tool for both red and blue teaming exercises. However, their tools is only designed for the macOS operating system.
	
	Miller et al. \cite{miller2018automated} state that adversary emulation tools can be a cost-effective alternative to red teaming events. They mention several benefits of adversary emulation tools, such as providing defenders with a view on their network from the point of an attacker, identifying weaknesses or misconfigurations, testing of deployed security measures, and providing empirical evidence for a defensive blue team. The authors state that adversary emulation tools should be (i) intelligent, i.e., select and chain actions similar to an actual adversary, (ii) usable, i.e., require low overhead for utilization, (iii) realistic, i.e., execute attack techniques in such a way as they occur in the real world, and (iv) modular, i.e., allow users to customize techniques and create new attack procedures.
	
	These and similar requirements are at the core of scientific reviews and surveys on the topic of adversary emulation. Zilberman et al. \cite{zilberman2020sok} provide one of the most comprehensive surveys, comprising eleven tools and 45 criteria. One of the focus points of their study is to assess how many attack techniques from MITRE ATT\&CK are covered by each tool. In addition, they evaluate compatibility with operating systems, prerequisites for installation and running the tools, ease-of-use, documentation, and many technical features such as logging, cleanup, and the creation of custom attack scenarios. The main difference to our work is that we conduct a study with domain experts to better understand and weight the importance of each of these requirements. Contrary to their work, we also put less focus on the coverage of attack techniques and instead include more fine-granular questions on other aspects, which we outline in Sect. \ref{methodology}.
	
	Other studies on adversary emulation tools have a more narrow focus. Orbinato et al. \cite{orbinato2024laccolith} compare multiple adversary emulation tools with respect to their ability to evade detection by antivirus. Their results suggest that their own approach, which involves an agent that resides in the kernel, is more reliable in evading detection than most other tools, which require to manually configure exceptions in detection tools. The issue of detection is also studied by Elgh et al. \cite{elgh2022comparison}, who analyze the number of Sysmon log events generated on a Windows target host while four adversary emulation tools are actively used to attack the machine. Since log events are a main source for intrusion detection, it is essential that automated attack simulations do not produce significantly more or more severe events than manual execution of attacks. However, their comparative study suggest that most tools are too noisy to represent realistic attacks from actual advanced persistence threats. Stockenreitner et al. \cite{stock2022faehigkeitsanalyse} compare four tools with respect to their coverage of MITRE ATT\&CK techniques when it comes to predefined attack procedures targeting Windows and Active Directory. The study shows that most tools only have limited capabilities in Windows environments. Other than these works, our survey is not targeted at specific properties of adversary emulation tools but instead relies on user studies to identify important aspects.
	
	\section{Methodology} \label{methodology}
	
	This section outlines our research methodology, including strategies to select adversary emulation tools and create questionnaires for tool evaluations and user requirements analysis.
	
	\subsection{Overview}
	
	\begin{figure}
		\centering
		\includegraphics[width=1\columnwidth]{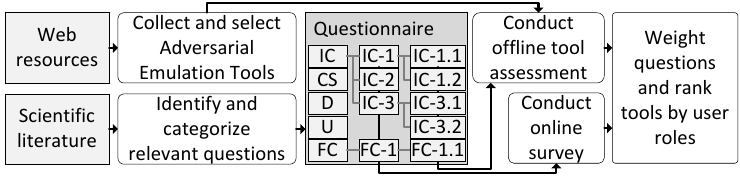}
		\caption{Overview of the methodology of our research.}
		\label{fig:meth}
		\vspace{-7pt}
	\end{figure} 
	
	Figure \ref{fig:meth} depicts our methodology as a flowchart. Initially, we select a set of adversary emulation tools for our study by conducting a web search for well-known open-source tools (cf. Sect. \ref{toolselection}). Based on literature research and insights gained from related studies, we then assemble a questionnaire that enables assessment of these tools. Thereby, we consider the following five categories of questions: 
	
	\begin{itemize}
		\item \textit{Installation \& Configuration (IC)}. Preparing systems and configurations in such a way to enable effective utilization of certain tools can often be time-consuming and pose a burden to analysts. This category covers several sub-categories, including compatibility of the tool with different operating systems, requirements on third-party tools, and tool configurability.
		\item \textit{Community \& Support (CS)}. An active community behind an open-source tool is highly valuable when questions or issues emerge during its setup or utilization as they provide rapid support and may even fix or extend the tool in future releases following user feedback. This category of questions aims to assess aspects regarding the popularity of the repository as well as the activity of involved developers.
		\item \textit{Documentation (D)}. The availability and extensiveness of documentation is an important source of information for analysts who deploy and use the tool. This category covers metrics such as the comprehensiveness, currency, comprehensibility, and standardization of tool documentation.
		\item \textit{Usability (U)}. Well-designed user interfaces can vastly improve user experience when interacting with a tool. Questions in this category thus focus on assessing the intuitiveness, appearance, and customizability of available user interfaces, such as graphical user interfaces or command line interfaces.
		\item \textit{Features \& Capabilities (FC)}. While some core functionalities required for adversary emulation are shared among all tools, some of them come with additional features and capabilities that are essential for certain use-cases. The sub-categories in this group of question address relevant technical aspects such as the overall workflow, restoring target systems after attacks, reporting and logging, chaining of attacks, attack automation, coverage of diverse attack techniques, customization and scripting of attacks, and flexibility of attack execution.
	\end{itemize}
	
	The central part of Fig. \ref{fig:meth} visualizes the structure of the questionnaire; specifically, each category of questions (e.g., \textit{IC}) are divided into one or more sub-categories (e.g., \textit{IC-1}), which in turn comprise one or more questions (e.g., \textit{IC-1.1}). We process the questionnaire in a two-step procedure. First, we conduct an offline evaluation and assess how each question is fulfilled by every tool through experimentation (cf. Sect. \ref{offline}). Second, we carry out an online survey where we ask stakeholders about their preferences and requirements on adversary emulation tools (cf. Sect. \ref{online}). As a final part of our study, we combine the results from the offline evaluation and online survey to assign a single score to each tool for the purpose of ranking (cf. Sect. \ref{ranking}). In the following sections, we describe the steps of our methodology in more detail.
	
	\subsection{Selection of adversary emulation tools} \label{toolselection}
	
	For our comparative study, we aim to select a manageable set of well-known open-source tools for adversary emulation. To ensure that the identified tools are established technologies of practical relevance, we scan through articles comparing various penetration testing tools\footnote{\url{https://tinyurl.com/wayback-2021/pentestit.com/adversary-emulation-tools-list/}} and validate with curated lists of threat hunting software\footnote{\url{https://github.com/0x4D31/awesome-threat-detection}} and tools analyzed in scientific state of the art \cite{zilberman2020sok, elgh2022comparison, orbinato2024laccolith}. 
	Thereby, we only include tools that are suitable for attack simulation and exclude tools for other types of security analytics, such as NSA Unfetter\footnote{\url{https://nsacyber.github.io/unfetter/}} that enables vulnerability identification but does not support attack execution. Moreover, we focus on stand-alone tools and exclude frameworks that only build on top of such tools, such as MATE\footnote{\url{https://github.com/fugawi/mate}}, Purple Team ATT\&CK\footnote{\url{https://github.com/praetorian-inc/purple-team-attack-automation}}, Splunk Attack Range\footnote{\url{https://github.com/splunk/attack_range}}, or ATT\&CK Simulator\footnote{\url{https://github.com/timfrazier1/AdversarySimulation}}. 
	
	Eventually, we end up with the following nine open-source tools for adversary emulation (sorted alphabetically): ATTPwn\footnote{\url{https://github.com/Telefonica/ATTPwn}}, Atomic Red Team\footnote{\url{https://github.com/redcanaryco/atomic-red-team}}, APTSimulator\footnote{\url{https://github.com/NextronSystems/APTSimulator}}, MITRE Caldera\footnote{\url{https://github.com/mitre/caldera}}, DumpsterFire\footnote{\url{https://github.com/TryCatchHCF/DumpsterFire}}, Infection Monkey\footnote{\url{https://github.com/guardicore/monkey}}, Invoke Adversary\footnote{\url{https://github.com/CyberMonitor/Invoke-Adversary}}, Metasploit\footnote{\url{https://github.com/rapid7/metasploit-framework}}, and Purplesharp\footnote{\url{https://github.com/mvelazc0/PurpleSharp}}. We provide descriptions for each of these tools in Sect. \ref{tools}. We point out that several commercial products for adversary emulation exist but are excluded from this study for licensing issues.
	
	\subsection{Offline Tool Assessment} \label{offline}
	
	This section describes how we assembled and subsequently answered the questions from the questionnaire. We used the study by Zilberman et al. \cite{zilberman2020sok} as a guide to identify and validate relevant questions for most categories. Specifically, we use six questions on prerequisites and compatibility from that study in category \textit{IC}, six questions about comprehensiveness of documentation for category \textit{D}, six questions about availability of core functionalities over various interfaces (e.g., attack execution and configuration over graphical or command line interface) for category \textit{U}, and 29 questions on technical aspects for category \textit{FC}. Based on the work by Joy et al. \cite{joy2018performance}, who analyze open-source projects and validate that their performance can be assessed through public metrics such as number of forks, we formulate six questions for category \textit{IC}. Aversano et al. \cite{aversano2017evaluating} propose quality indicators for documentation of open-source software, which we use to formulate eight questions in category \textit{D}. Richter et al. \cite{richter1997kriterien}, state several criteria for usability in human-machine-systems, which we use to formulate 13 additional questions for category \textit{U}. Based on our own experience, we pose five questions that deal with attack chaining, blue teaming, and tool workflow in category \textit{FC} and one question that checks conformity to the \textit{IEEE Standard for User Documentation} \cite{5712775} in category \textit{D}.
	
	Eventually, we end up with 80 questions (see Appendix \ref{appendix} for a list of all questions) that we use to evaluate each of the selected adversary emulation tools. We answer questions from categories \textit{CS} and \textit{D} by reviewing the documentation and resources provided in the respective repository of the tool. Regarding questions from categories \textit{IC}, \textit{U}, and \textit{FC}, we proceed by installing each adversary emulation tool in a virtual environment and testing the features according to the questions, e.g., by executing attack cases. The setup comprises a Kali-Linux\footnote{\url{https://www.kali.org/}} machine (Version 2024 with 4 GB RAM and 60 GB disk space) that acts as a command-and-control server, two Metasploitable3\footnote{\url{https://github.com/rapid7/metasploitable3}} machines (Ubuntu 14.04 with 4 GB RAM and 20 GB disk space as well as Windows Server 2008 with 4 GB RAM and 40 GB disk space) that act as target systems, and two Windows machines (Versions 10 and 11, each with 8 GB RAM and 80 GB disk space) that act as both command-and-control servers and target systems. We also cross-check the documentation to ensure that no functions supported by the tools are missed or misapplied.
	
	Similar to the survey conducted by Zilberman et al. \cite{zilberman2020sok}, our assessment scheme consists of a four-point-scale, where we assign 0 points if the question is not fulfilled, 1 point for partial fulfillment, 2 points when the question is mostly fulfilled, and 3 points if it is entirely fulfilled. For questions that are answered with yes or no, we assign 3 and 0 points respectively. Since category \textit{CS} comprises quantitative questions, we assign 3 points if the retrieved value is in the top 25\% of all gathered values for that question, 1 point if it is in the bottom 25\%, and 2 points if it is in between.
	
	\subsection{Online Survey of User Requirements} \label{online}
	
	The second part of our study aims to assess which of the aforementioned functions and properties of adversary emulation tools are most important for stakeholders to enable weighting and ranking of tools. To this end we invite domain experts to an online survey and ask them to assign relevance scores to each of the 30 sub-categories of questions. We opted for sub-categories rather than the entire questionnaire (c.f. Sect. \ref{offline}) to reduce the number of questions from 80 to 30 and avoid asking for many technical specifics. One of them is a general statement \textit{G-1: Tool is available for free} that we add even though it is not related to any category, but only used to assess the preference of freely available tools in contrast to commercial products. We design the survey following the work by Achimugu et al. \cite{achimugu2014adaptive}, who outline the Fuzzy Multi-Criteria Decision-Making (FMCDM) method for prioritization of software requirements. In particular, participants can rate the importance of each sub-category on a five point scale ranging from \textit{Not Important} to \textit{Very Important}, also including \textit{No answer} as an additional option.
	
	Beside these questions, we ask participants to state their professional experience, where we differentiate between categories \textit{Low} (less than 3 years), \textit{Medium} (3 to 6 years), and \textit{High} (more than 6 years), as well as their expertise with adversary emulation tools, where we differentiate between categories \textit{Low} (not familiar), \textit{Medium} (somewhat familiar), and \textit{High} (very familiar). In addition, we ask participants about their current job position to identify differences in preferences across groups of users. Specifically, we group all participants to one of the following roles: \textit{Leaders} (i.e., executives and managers), \textit{Security Architects} (e.g., security engineers and system administrators), \textit{Security Consultants}, \textit{Security Analysts}, and \textit{Security Researchers}. 
	
	The survey was hosted on a web platform from January 15, 2024, to March 15, 2024, and shared through a link that we posted on cyber security mailing lists and within project consortia to attract participants. At the beginning of the survey, all participants were informed about the purpose of the survey and consented that their responses will be published anonymously as part of this study.
	
	\subsection{Feature Weights and Tool Ranking} \label{ranking}
	
	We combine the results obtained from our offline tool evaluation with the responses from our online survey to weight properties of adversary emulation tools and create rankings for certain user groups. To this end we map the five point scale of user importance ratings to numeric weights as follows: \textit{Not Important} has a weight of $0.5$, \textit{Rather Unimportant} has a weight of $0.75$, \textit{Neutral} has a weight of $1$, \textit{Important} has a weight of $1.25$, and \textit{Very Important} has a weight of $1.5$. For each user role, we then compute the weight of each sub-category of questions as the average of all responses from participants that belong to that role. In the following, we denote the average weights as $\overline{w}_{subcat(q), role}$, where $subcat(q)$ is the sub-category of question $q$.
	
	To compute an overall score for one of the reviewed adversary emulation tools, we iterate over the list of questions $Q_{tool}$ that we evaluated in the offline tool assessment and multiply the score $score(q) \in \left\lbrace 0, 1, 2, 3 \right\rbrace $ of question $q$ with the weight corresponding to $subcat(q)$. As depicted in Eq. \ref{pw}, we compute the final score $s_{tool, role} \in \left[0, 1 \right] $ by normalizing with $3 \cdot \left| Q_{tool} \right|$, since $3$ is the maximum score for each question. 
	
	\begin{equation} \label{pw}
		s_{tool, role} = \frac{1}{3 \cdot \left| Q_{tool} \right| } \sum_{q \in Q_{tool}} score(q) \cdot \overline{w}_{subcat(q), role}
	\end{equation}
	
	Note that the total highest achievable number of points is different for each tool, because there are two questions in sub-category \textit{FC} that are only applicable to tools that rely on agents. Accordingly, these questions are not evaluated for tools without agents and thus do not contribute to their scores.
	
	\section{Adversary Emulation Tools} \label{tools}
	
	\begin{figure}
		\centering
		\includegraphics[width=1\columnwidth]{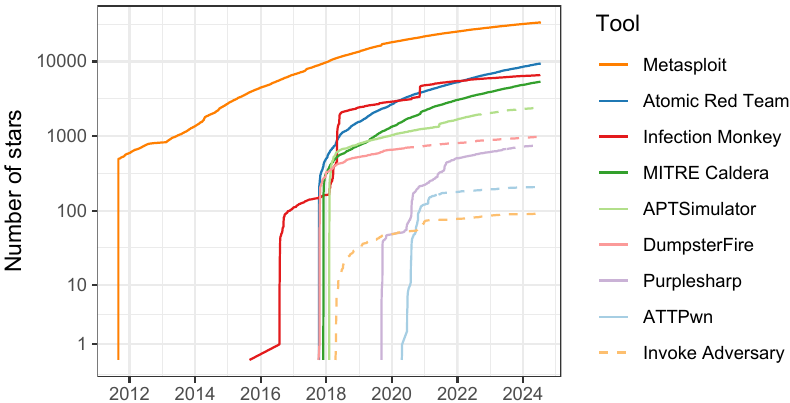}
		\caption{Progression of received stars for the respective GitHub repositories of each tool. Transition from solid to dashed lines indicate the last commit made to the repository.}
		\label{fig:stars}
		\vspace{-7pt}
	\end{figure} 
	
	\begin{table}[]
		\small 
		\caption{Operating system support and requirement to install agents on target systems for attack executions}
		\label{tab:tools}
		\begin{tabular}{lcccc}
			& \multicolumn{3}{c}{\textbf{Supported Operating Systems}} &  \\ \cline{2-4}
			\textbf{Tool} & \textbf{Windows} & \textbf{Linux} & \textbf{MacOS} & \textbf{Agents}  \\ \hline \hline
			ATTPwn& \checkmark & &  & \checkmark  \\ \hline
			Atomic Red Team & \checkmark & \checkmark & \checkmark &    \\ \hline
			APTSimulator & \checkmark & &  &    \\ \hline
			MITRE Caldera & & \checkmark & \checkmark & \checkmark  \\ \hline
			DumpsterFire & \checkmark & \checkmark & \checkmark &   \\ \hline
			Infection Monkey & \checkmark & \checkmark &  & \checkmark  \\ \hline
			Invoke Adversary  & \checkmark & &  &    \\ \hline
			Metasploit & \checkmark & \checkmark & \checkmark & $\sim$    \\ \hline
			Purplesharp & \checkmark & & & \checkmark  \\ \hline
		\end{tabular}
		\vspace{-7pt}
	\end{table}
	
	This section briefly describes each of the selected adversary emulation tools and highlights some of their properties that differentiate them from others. Since all tools are available on GitHub, we provide a visual summary of the age and popularity of the respective repositories in Fig. \ref{fig:stars}. The change from solid to dashed lines indicates the point in time where the most recent commit was made to the repository. As visible in the plot, the most popular repositories are still maintained while some of the other repositories have not been updated for years and can be considered as discontinued by developers. As visible in the plot, most repositories receive a significant amount of stars in the first few weeks after release and then continue to slowly grow in popularity over time.
	
	We also summarize the compatibility of each tool with operating systems as well as the need to install an agent on the target system in Table \ref{tab:tools}. The overview shows that Windows is the most widely supported operating system, but popular repositories (according to Fig. \ref{fig:stars}) also support alternative operating systems. In the following, we go through each tool in alphabetical order.
	
	\textbf{ATTPwn} emulates threats following the schema defined by MITRE ATT\&CK and specifically covers many techniques from the \textit{Defense Evasion} and \textit{Discovery} stages. The tool focuses on Windows since installation requires Powershell 3.0 or above and most available attack techniques target Windows machines, even though procedures for other operating systems are available. Running the attacks requires that an agent is downloaded from the command-and-control server to the target system, which requires to add exceptions to firewalls and antivirus. The graphical user interface is simple and well-arranged, but does not provide any support or feedback for users. Nonetheless, it allows to review and modify attack scripts, even for attack chains. The most striking difference to other tools is the lack of documentation that is limited to a short readme file and some video tutorials. 
	
	\textbf{Atomic Red Team} comes with a large library of predefined attack procedures, which is one of its key features in comparison to other tools. Moreover, it includes scripts in markdown and YAML format that support efficient setup of reproducible and portable test environments. Both Atomic Red Team and its attack-executor Invoke-Atomic are simple to install on all common operating systems. The framework provides a graphical user interface that allows efficient generation of attack procedures as well as a command line interface to execute attack procedures from the library. Knowledge of the Python programming language is required to modify, combine, or create entirely new attack steps. Due to the straightforward design of the tools, the comparatively short documentation is sufficient to understand and use all available functions.
	
	\textbf{APTSimulator} is a Windows batch script that emulates a compromised Windows system without the need for an agent. Due to the straightforward nature of the tool, the short documentation in form of a readme file is sufficient to retrieve and run the script. However, we noticed during our experiments that it is necessary to apply exceptions to antivirus systems in Windows in order to run the script. Attack procedures are organized in a schema similar to MITRE ATT\&CK and can be executed in batches; in particular, all available attacks can be run in a sequence. Creation of new attack cases is not supported through the command line interface, but requires to modify existing batch scripts or create new ones.
	
	\textbf{MITRE Caldera} is designed to support cyber security teams with autonomous and reproducible attack simulations that are useful for testing of detection, analysis, and response capabilities. Interestingly, it is the only tool that is not compatible with Windows, since the command-and-control server needs to be installed either on Linux or MacOS. The tool excels in terms of quality of documentation, which contains comprehensive explanations of all relevant components, tutorials for execution of attacks, and even a chapter for developers. In addition, MITRE Caldera offers interactive training to learn the basic features of the tool. 
	
	The tool also stands out from the others due to the combination of high usability and extensive number of features. The graphical user interface is accessible through a browser and provides configuration options, an enumeration of available attack procedures, and an overview of statuses for agents as well as currently ongoing attacks. Attack executions may be stopped or interrupted to add new procedures. To run the attack simulations, agents must be installed on the target systems, which can also be carried out through the graphical user interface. Similar to most other tools, agent installation requires exceptions in firewall and antivirus on target systems. 
	
	Beside preconfigured attack procedures, MITRE Caldera integrates procedures from other frameworks such as Metasploit. Users have the possibility to create new attack procedures as well as to modify existing ones, arrange them in chains for parallel or sequential execution, and specify cleanup functions that are executed after the fact. The tool also provides comprehensive logging and reporting functionalities, which support output in JSON, CSV, text, or PDF files that even contain visualizations of the compromised infrastructure.
	
	\textbf{DumpsterFire} is a platform-independent tool based on the Python programming language that is designed to generate repeatable, delayed, and distributed security events. Users may change existing and add new attack procedures as Python scripts. Other than for most of the reviewed tools, the predefined attack procedures are not categorized following any model such as MITRE ATT\&CK and include activities without direct security implications, such as opening a browser and playing videos. Moreover, the tool lacks some features present in other tools, such as cleanup functions to restore target systems or the option to stop currently ongoing attack procedures. DumpsterFire only provides a command line interface with some basic help pages as well as a readme file. While the console output itself can be regarded as a basic report, the tool does not produce any logs or output files. Given that there is only a single contributor to the repository and the most recent commit dates to the year 2020, we assume that the project has been discontinued.
	
	\textbf{Infection Monkey} is designed to test security solutions. The tool relies on a worm-like agent on the target system that scans the network, executes attack procedures, and simulates lateral movement. Both agent and command-and-control server are compatible with Windows and Linux. The command-and-control server is a web server providing a graphical user interface that allows users to start and configure attack procedures or store them as reusable templates. Moreover, it provides a so-called Infectionmap showing compromised infrastructures as well as information about target hosts. As for most tools, firewall and antivirus block both the agent and server and must be disabled. Infection Monkey generates detailed log files for executed procedures and allows to generate a comprehensive report. The documentation covers all functions of the tool, but does not explain how users can create their own attack procedures, which requires programming skills and cannot be accomplished in the graphical user interface. 
	
	\textbf{Invoke Adversary} is a Powershell script useful to assess security tools. Similar to APTSimulator, it is straightforward to get started since there is no installation or deployment of agents required; running the script is sufficient. The tool comprises a menu in the command line interface that allows to select and execute attack procedures. As there are no predefined techniques for lateral movement available, the tool is primarily useful to assess endpoint detection systems. To generate new attack procedures, the script itself must be adapted accordingly. Logging of executed procedures is limited to the command line as no log files or reports are stored. Note that exceptions for antivirus are necessary to run the tool. The community and support behind the tool appears to be comparatively limited, and there have not been any updates for the tool since its initial release.
	
	\textbf{Metasploit} is the oldest and by far the most popular penetration testing framework among all reviewed tools. It is backed up by a large and active community of users and developers and mainly used to detect and assess vulnerabilities of systems and networks. In contrast to other tools from our study that either require an agent on the target system or not, Metasploit finds a middle ground as its Meterpreter allows to establish an interactive shell, execute commands, and reload payloads on the target similar to an agent. Attack procedures may be represented as reusable templates and chained together; however, to generate new procedures, users need to define them in scripts in the Ruby programming language. 
	
	Logging in Metasploit is configurable for different log levels. While Metasploit is a command line tool, we point out that there is also a commercial version of Metasploit that provides a graphical user interface but is excluded from our study as we only focus on open-source tools. Metasploit comes with 875 pages of documentation describing all modules in detail and providing guidance for developers; this is the most comprehensive documentation of all reviewed tools.
	
	\textbf{Purplesharp} is written in C-Sharp and designed to carry out attacks on Windows Active Directory. On the target host, three agents are used for reconnaissance, attack execution, and orchestration of attacks. Antivirus tools need to be disabled to run the agents. The tool is started through the Windows command line, but also provides a graphical user interface in a browser that can be used to configure available attack procedures. To generate new attack procedures, however, programming skills are necessary. In general, the simple nature of the tool enables automation and chaining of attack procedures as well as inclusion of cleanup functions through playbooks in JSON format. The documentation is sufficient to install the tools and carry out predefined attack procedures. Purplesharp also produces log files containing relevant information from executed attack procedures.
	
	\section{Evaluation} \label{evaluation}
	
	This section contains our evaluation results. We first present a qualitative comparison of adversary emulation tools based on our offline evaluation. Subsequently, we provide an overview of the prioritized user requirements that we obtain from our online survey. Finally, using the weighted features, we rank adversary emulation tools for different user roles.
	
	\subsection{Technical Comparison} \label{qualitative}
	
	\begin{table*}[]
		\caption{Results of the offline evaluation}
		\label{tab:res}
		\begin{tabular}{lcccccccccccccccc}
			& \multicolumn{3}{c}{\makecell[c]{Installation \& \\ Configuration (IC)}}                            & \multicolumn{3}{c}{\makecell[c]{Community \& \\ Support (CS)}}                            & \multicolumn{3}{c}{Documentation (D)}                            & \multicolumn{3}{c}{Usability (U)}                            & \multicolumn{3}{c}{\makecell[c]{Features \& \\ Capabilities (FC)}}                            \\ \cline{2-16}
			& \multicolumn{1}{c}{\makecell[c]{Pts. \\ abs.}} & \multicolumn{1}{c}{\makecell[c]{Pts. \\ max.}} & \makecell[c]{Pts. \\ rel. (\%)} & \multicolumn{1}{c}{\makecell[c]{Pts. \\ abs. }} & \multicolumn{1}{c}{\makecell[c]{Pts. \\ max.}} & \makecell[c]{Pts. \\ rel. (\%)} & \multicolumn{1}{c}{\makecell[c]{Pts. \\ abs.}} & \multicolumn{1}{c}{\makecell[c]{Pts. \\ max.}} & \makecell[c]{Pts. \\ rel. (\%)} & \multicolumn{1}{c}{\makecell[c]{Pts. \\ abs.}} & \multicolumn{1}{c}{\makecell[c]{Pts. \\ max.}} & \makecell[c]{Pts. \\ rel. (\%)} & \multicolumn{1}{c}{\makecell[c]{Pts. \\ abs.}} & \multicolumn{1}{c}{\makecell[c]{Pts. \\ max.}} & \makecell[c]{Pts. \\ rel. (\%)} \\ \hline
			\multicolumn{1}{l}{ATTPwn} & \multicolumn{1}{c}{13} & \multicolumn{1}{c}{18} & 72.2 & \multicolumn{1}{c}{9} & \multicolumn{1}{c}{18} & 50.0 & \multicolumn{1}{c}{0} & \multicolumn{1}{c}{42} & 0.0 & \multicolumn{1}{c}{19} & \multicolumn{1}{c}{57} & 33.3 & \multicolumn{1}{c}{65} & \multicolumn{1}{c}{102} & 63.7 \\ \hline
			\multicolumn{1}{l}{Atomic Red Team} & \multicolumn{1}{c}{18} & \multicolumn{1}{c}{18} & \textbf{100.0} & \multicolumn{1}{c}{17} & \multicolumn{1}{c}{18} & 94.4 & \multicolumn{1}{c}{35} & \multicolumn{1}{c}{42} & 83.3 & \multicolumn{1}{c}{17} & \multicolumn{1}{c}{57} & 29.8 & \multicolumn{1}{c}{74} & \multicolumn{1}{c}{96} & 77.1 \\ \hline
			\multicolumn{1}{l}{APTSimulator} & \multicolumn{1}{c}{13} & \multicolumn{1}{c}{18} & 72.2 & \multicolumn{1}{c}{10} & \multicolumn{1}{c}{18} & 55.6 & \multicolumn{1}{c}{24} & \multicolumn{1}{c}{42} & 57.1 & \multicolumn{1}{c}{19} & \multicolumn{1}{c}{57} & 33.3 & \multicolumn{1}{c}{42} & \multicolumn{1}{c}{96} & 43.8 \\ \hline
			\multicolumn{1}{l}{MITRE Caldera} & \multicolumn{1}{c}{17} & \multicolumn{1}{c}{18} & 94.4 & \multicolumn{1}{c}{13} & \multicolumn{1}{c}{18} & 72.2 & \multicolumn{1}{c}{36} & \multicolumn{1}{c}{42} & \textbf{85.7} & \multicolumn{1}{c}{36} & \multicolumn{1}{c}{57} & \textbf{63.2} & \multicolumn{1}{c}{84} & \multicolumn{1}{c}{102} & \textbf{82.4} \\ \hline
			\multicolumn{1}{l}{DumpsterFire} & \multicolumn{1}{c}{15} & \multicolumn{1}{c}{18} & 83.3 & \multicolumn{1}{c}{11} & \multicolumn{1}{c}{18} & 61.1 & \multicolumn{1}{c}{17} & \multicolumn{1}{c}{42} & 40.5 & \multicolumn{1}{c}{24} & \multicolumn{1}{c}{57} & 42.1 & \multicolumn{1}{c}{36} & \multicolumn{1}{c}{96} & 37.5 \\ \hline
			\multicolumn{1}{l}{Infection Monkey} & \multicolumn{1}{c}{14} & \multicolumn{1}{c}{18} & 77.8 & \multicolumn{1}{c}{12} & \multicolumn{1}{c}{18} & 66.7 & \multicolumn{1}{c}{26} & \multicolumn{1}{c}{42} & 61.9 & \multicolumn{1}{c}{31} & \multicolumn{1}{c}{57} & 54.4 & \multicolumn{1}{c}{43} & \multicolumn{1}{c}{102} & 42.2 \\ \hline
			\multicolumn{1}{l}{Invoke Adversary} & \multicolumn{1}{c}{13} & \multicolumn{1}{c}{18} & 72.2 & \multicolumn{1}{c}{7} & \multicolumn{1}{c}{18} & 38.9 & \multicolumn{1}{c}{12} & \multicolumn{1}{c}{42} & 28.6 & \multicolumn{1}{c}{13} & \multicolumn{1}{c}{57} & 22.8 & \multicolumn{1}{c}{24} & \multicolumn{1}{c}{96} & 25.0 \\ \hline
			\multicolumn{1}{l}{Metasploit} & \multicolumn{1}{c}{15} & \multicolumn{1}{c}{18} & 83.3 & \multicolumn{1}{c}{18} & \multicolumn{1}{c}{18} & \textbf{100.0} & \multicolumn{1}{c}{33} & \multicolumn{1}{c}{42} & 78.6 & \multicolumn{1}{c}{21} & \multicolumn{1}{c}{57} & 36.8 & \multicolumn{1}{c}{77} & \multicolumn{1}{c}{96} & 80.2 \\ \hline
			\multicolumn{1}{l}{Purplesharp} & \multicolumn{1}{c}{13} & \multicolumn{1}{c}{18} & 72.2 & \multicolumn{1}{c}{9} & \multicolumn{1}{c}{18} & 50.0 & \multicolumn{1}{c}{21} & \multicolumn{1}{c}{42} & 50.0 & \multicolumn{1}{c}{16} & \multicolumn{1}{c}{57} & 28.1 & \multicolumn{1}{c}{43} & \multicolumn{1}{c}{102} & 42.2 \\ \hline
		\end{tabular}
		\vspace{-7pt}
	\end{table*}
	
	We conducted the offline evaluation of adversary emulation tools according to our outline from Sect. \ref{offline}. Table \ref{tab:res} summarizes the results of the evaluation. For each category of questions, we state the total number of points achieved by an adversary emulation tool (\textit{Pts. abs.}), the maximum number of achievable points (\textit{Pts. max}), and the relative number of achieved points (\textit{Pts rel. (\%)}). Note that the maximum number of points is different in each category but also within category \textit{FC} since some questions only apply to agent-based tools and are not evaluated for tools that do not make use of agents. 
	
	In category \textit{IC}, most tools achieve relatively high points, with the minimum score of $72.2\%$ being achieved by four tools. This indicates that technical obstructions that make it difficult to get started with a tool are quite successfully kept to a minimum. In general, most points in this category are deducted for incompatibility with common operating systems (\textit{IC-1.1}) and required changes to security settings (\textit{IC-1.2}). In particular, running the tools often requires to add exceptions in the firewall and antivirus of systems where the tool itself or its agents are deployed. Atomic Red Team is the only tool to achieve the highest possible score of $18$ points as it is compatible with all three considered operating systems and does not rely on agents. MITRE Caldera, despite relying on agents, notably achieves the second best score with $17$ points.
	
	Given that Metasploit is by far the most widely used tool for adversary emulation, it is not surprising that it achieves the highest possible score of $18$ points in category \textit{CS}. For comparison, even though Atomic Red Team achieves $17$ points in that category, Metasploit has around three times as many contributors (\textit{CS-1.2}) and five times as many forks (\textit{CS-1.1}) as Atomic Red Team. Regarding category \textit{D}, the table shows that MITRE Caldera yields the highest score and is closely followed by Atomic Red Team and Metasploit, while many other tools fall behind. Our analysis reveals that most documentations suffer from low readability scores (\textit{D-4.1}) and leave out descriptions on how to interpret the output of executed procedures (\textit{D-1.6}), generate custom procedures (\textit{D-1.4}), or arrange procedures as chains (\textit{D-1.5}). Captions of figures are also missing in almost all documentations (\textit{D-5.1}).
	
	Category \textit{U} turned out to be a critical one, where many tools only yield comparatively few points. MITRE Caldera yields the highest score with $36$ out of $57$ points and is followed by Infection Monkey with $31$ points; other tools all achieve only around $20$ points. This is primarily caused by limitations of user interfaces (\textit{U-5.3}-\textit{U-5.6}) and low flexibility (\textit{U-7.2}), in particular, many essential functions are often only available through one interface (e.g., command line) but not through another (e.g., graphical user interface). Other common issues include lack of customizability of interfaces (\textit{U-6.1}) and limited guidance within the tool itself (\textit{U-3.1}, \textit{U-3.2}). One a positive note, most tools yield reproducible results (\textit{U-1.1}), are hardly affected by errors (\textit{U-1.3}), and their core functions are easy to access and efficient to execute (\textit{U-7.1}). 
	
	High diversity of results is also prevalent in category \textit{FC}, with MITRE Caldera ($84$ out of $102$ points), Metasploit ($77$ out of $96$ points), and Atomic Red Team ($74$ out of $96$ points) again forming the top. The main reasons for point reductions are issues with firewall and antivirus during operation (\textit{FC-2.2}, \textit{FC-2.3}, \textit{FC-2.6}, \textit{FC-2.7}), limited control of attack execution (\textit{FC-5.3}, \textit{FC-6.1}, \textit{FC-9.1}), unsuitable presentation of results of attack executions (\textit{FC-4.3}), and lack of blue teaming functionalities (\textit{FC-7.4}). 
	
	\begin{figure}
		\centering
		\includegraphics[width=1\columnwidth]{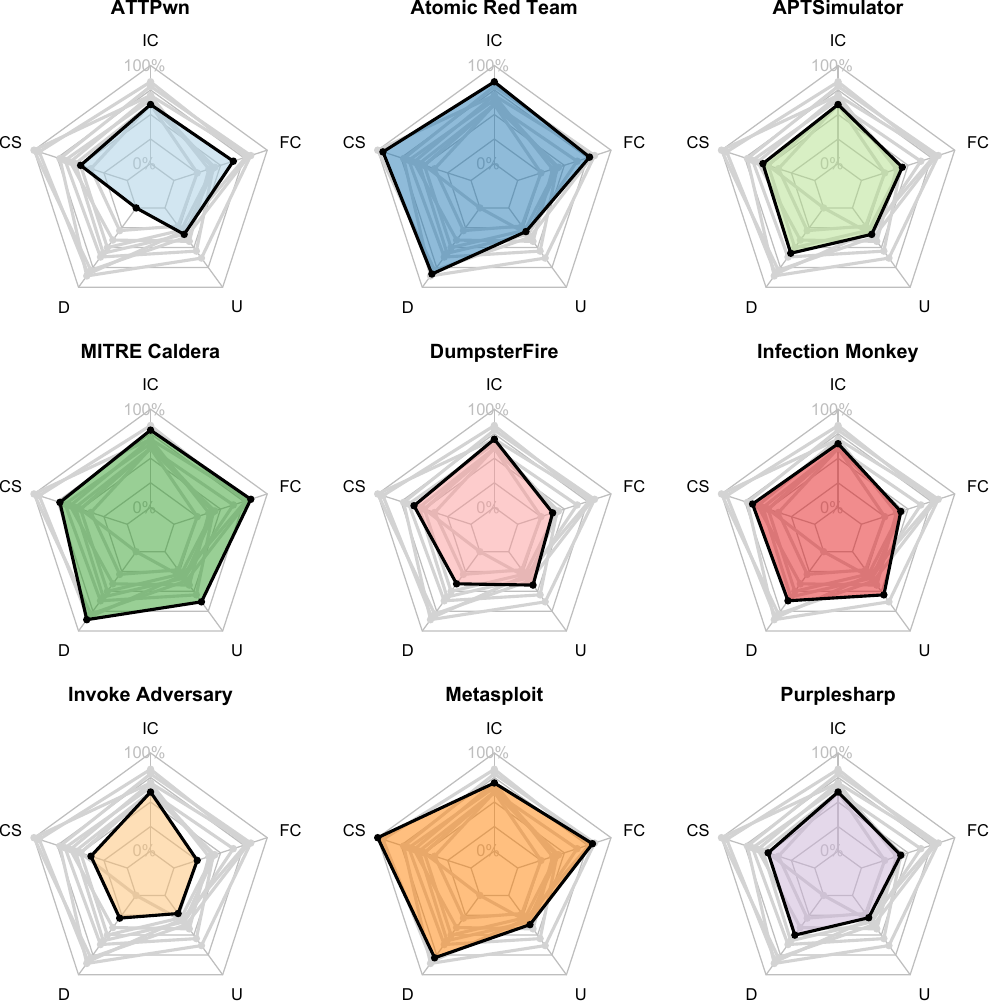}
		\caption{Radar plots of scores for categories Installation \& Configuration (IC), Community \& Support (CS), Documentation (D), Usability (U), and Features \& Capabilities (FC).}
		\label{fig:radar}
		\vspace{-7pt}
	\end{figure} 
	
	The radar plots displayed in Fig. \ref{fig:radar} provide a visual overview of the scores achieved by each tool. The plots show that Atomic Red Team and Metasploit have comparatively high and similar scores across all categories. MITRE Caldera exceeds them in terms of usability (category \textit{U}) but falls behind when it comes to support from the community (category \textit{CS}). The radar plots of DumpsterFire, APTSimulator, Infection Monkey, and Purplesharp have somewhat similar shapes that indicate medium fulfillment across all categories. ATTPwn would fall into a similar range, except that it yields the lowest possible score in category \textit{D}. Finally, Invoke Adversary appears to be at the lower end of the spectrum in most categories.
	
	\begin{figure}
		\centering
		\includegraphics[width=1\columnwidth]{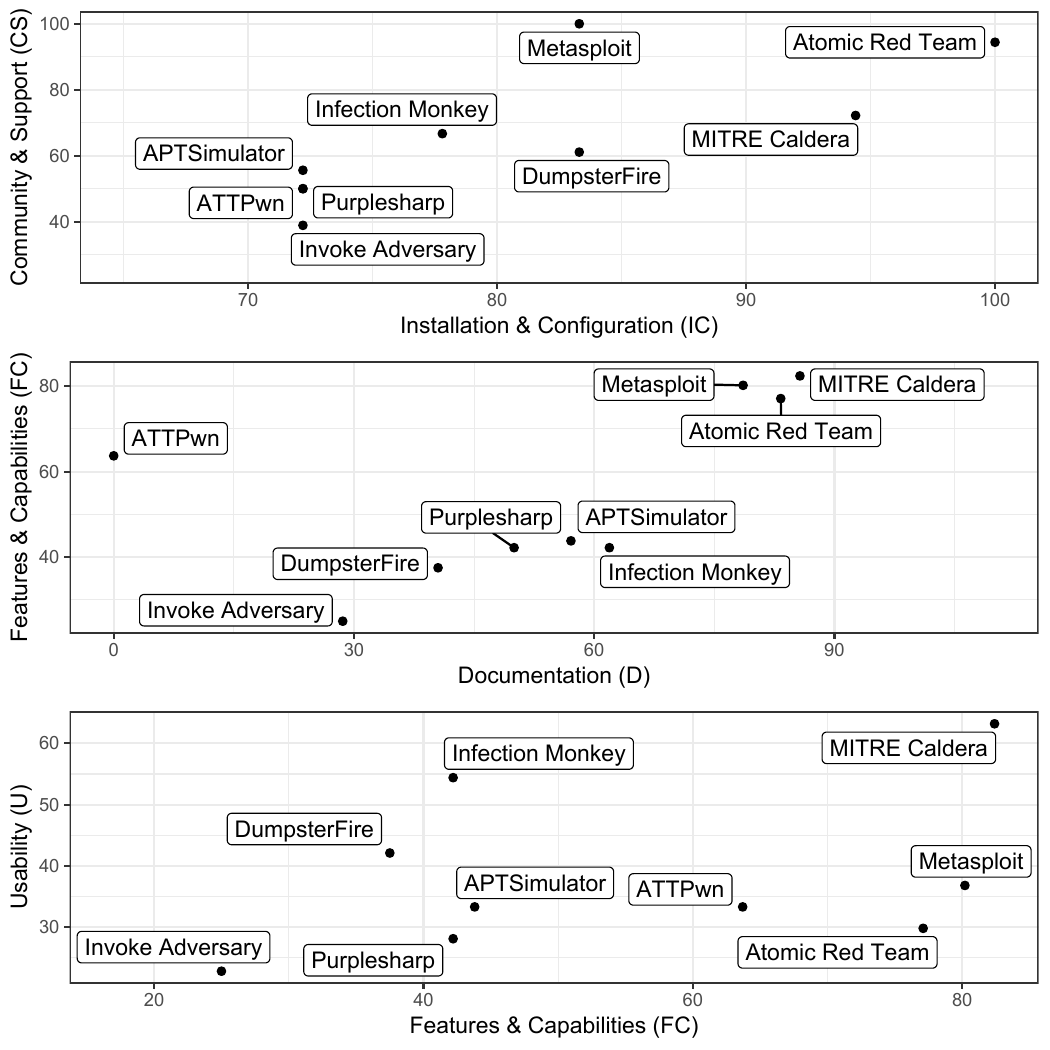}
		\caption{Pairwise comparison of category scores.}
		\label{fig:points}
		\vspace{-7pt}
	\end{figure} 
	
	Figure \ref{fig:points} visualizes the scores of pairs of categories for all tools. The top plot shows that Atomic Red Team is ahead of all other tools for categories \textit{CS} and \textit{IC}. The center plot compares categories \textit{FC} and \textit{D}, which reveals that MITRE Caldera, Atomic Red Team, and Metasploit form a group that outperforms all other tools. While AttPwn comes close in terms of \textit{FC}, it yields the lowest score in terms of \textit{D}. The bottom plot shows that MITRE Caldera is the only tool that combines high scores in categories \textit{U} and \textit{FC}. Metasploit and Atomic Red Team have comparable scores in category \textit{FC} but fall behind in terms of \textit{U}, and the exact opposite is the case for Infection Monkey. We provide a unified ranking that considers all categories in Sect. \ref{final}, but first present the results of our user requirement study in the following section that we subsequently use for weighting aforementioned criteria.
	
	\subsection{Online Survey Results} \label{online_results}
	
	This section contains the results of our online survey. We first provide some details on the participants of our survey and then present an overview of their responses.
	
	\subsubsection{Participants}
	
	\begin{table}[]
		\centering
		\footnotesize
		\setlength{\tabcolsep}{2pt}
		\caption{Online survey participants}
		\label{tab:participants}
		\begin{tabular}{llcc}
			\textbf{Role} & \textbf{Job Title} & \textbf{Experience} & \textbf{Expertise} \\ \hline \hline
			\multirow{5}{*}{Leaders} & Head of Cyber Security Research* & High & High \\ \cline{2-4} 
			& Head of Research* & High & High \\ \cline{2-4} 
			& Chief Information Security Officer  & High & Medium \\ \cline{2-4} 
			& Head of department & High & Medium \\ \cline{2-4} 
			& Head of department & High & High \\ \hline
			\multirow{4}{*}{\makecell[l]{Security \\ Architects}} & Senior Security Engineer & High & High \\ \cline{2-4} 
			& Enterprise Security Architect & High & Low \\ \cline{2-4} 
			& Security Engineer & High & High \\ \cline{2-4} 
			& DevSecOps Administrator & High & Low \\ \hline
			\multirow{3}{*}{\makecell[l]{Security \\ Consultants}} & Senior Security Consultant & High & High \\ \cline{2-4} 
			& Security Consultant & Low & Low \\ \cline{2-4} 
			& Security Consultant & High & Medium \\ \hline
			\multirow{4}{*}{\makecell[l]{Security \\ Analysts}} & SOC Analyst & Low & Medium \\ \cline{2-4} 
			& Senior Penetration Tester & Medium & High \\ \cline{2-4} 
			& Security Analyst & High & High \\ \cline{2-4} 
			& Security Specialist & Medium & Low \\ \hline
			\multirow{4}{*}{\makecell[l]{Security \\ Researchers}} & Head of Cyber Security Research* & High & High \\ \cline{2-4} 
			& Head of Research* & High & High \\ \cline{2-4} 
			& Scientist & Medium & Medium \\ \cline{2-4} 
			& Research Engineer & High & Medium \\ \hline
			\multirow{2}{*}{\makecell[l]{Unknown}} & N/A & N/A & High \\ \cline{2-4} 
			& N/A & High & Low \\ \hline
		\end{tabular}
	\end{table}
	
	Following our methodology from Sect. \ref{online} we published the survey over a period of two months, after which 20 domain experts fully completed the survey. Table \ref{tab:participants} provides an overview of the meta information we collected from all participants, specifically, their job title, job experience, and expertise with adversary emulation tools. After manually reviewing the job titles we grouped each participant into one of five roles, except for two participants (\textit{Head of Cyber Security Research} and \textit{Head of Research}) that fit into roles \textit{Leaders} and \textit{Security Researchers} and whose responses are thus counted in both groups. These participants are marked with an asterisk in Table \ref{tab:participants}. We opted for this setup to ensure that we base our estimations on sufficiently many users per role. Moreover, two participants did not state their job position and are thus assigned to the \textit{Unknown} role. Across all roles, most participants state high job experience, which we consider as an indicator for high-quality responses. On the other hand, participants indicate mixed expertise with adversary emulation tools, providing us with a diverse set of opinions on requirements. 
	
	\subsubsection{Responses} \label{responses}
	
	\begin{figure*}
		\centering
		\includegraphics[width=1\textwidth]{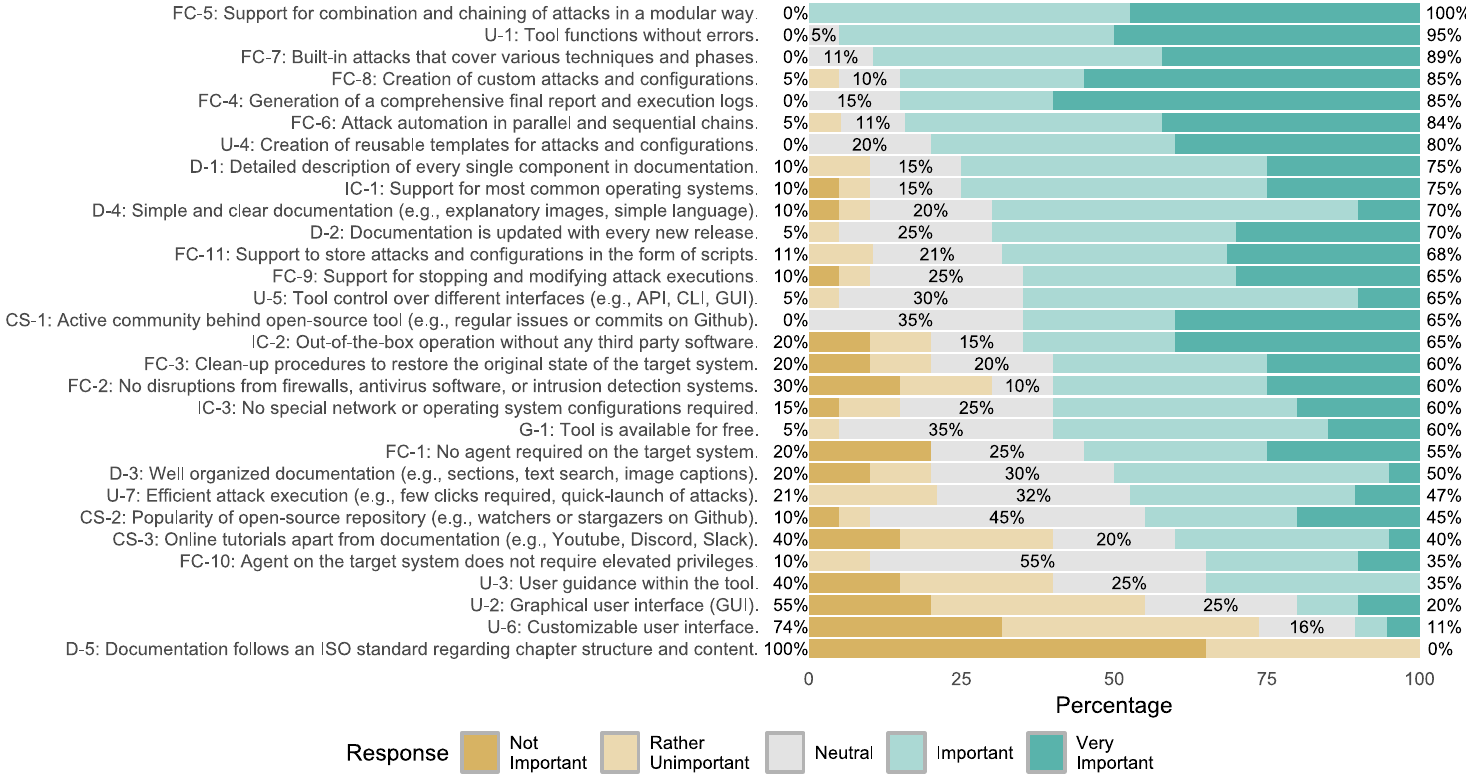}
		\caption{Likert scale of domain experts rating the importance of aspects of adversary emulation tools.}
		\label{fig:responses}
		\vspace{-7pt}
	\end{figure*} 
	
	We count the responses from all participants and visualize the resulting distributions in Fig. \ref{fig:responses}, which enumerates all $30$ sub-categories of questions. The sub-categories in the plot are sorted by the relative number of positive responses, i.e., categories mostly rated as \textit{Important} or \textit{Very Important} appear in the top of the plot. Our survey reveals that technical features from category \textit{FC} that primarily concern attack procedures form the most crucial requirements for adversary emulation tools, in particular, tools should come with a broad set of predefined attack procedures (\textit{FC-7}), but also enable to generate new attack procedures (\textit{FC-8}) and automate their execution in chains (\textit{FC-5}, \textit{FC-6}). Other relevant aspects are reporting (\textit{FC-4}) as well as stability (\textit{U-1}). The bottom part of the figure indicates that many aspects from categories \textit{D}, \textit{CS}, and \textit{U} are considered less relevant.
	
	We also investigate the responses with respect to roles, in particular, we counted the number of times participants of certain roles selected \textit{Important} and \textit{Very Important} for each category of questions. Our analysis suggests that \textit{Leaders} prioritize \textit{FC} and \textit{CS} over other categories, \textit{Security Architects} prioritize \textit{CS} and \textit{IC}, \textit{Security Consultants} prioritize \textit{CS} and \textit{FC}, and \textit{Security Analysts} prioritize \textit{IC} and \textit{FC}. Out of all roles, \textit{Security Researchers} assign the highest weight to \textit{FC} and the lowest weight to \textit{U}. In the following, we make use of these role-specific weights to compare and rank tools.
	
	\subsection{Weighted Results and Tool Ranking} \label{final}
	
	\begin{figure*}
		\centering
		\includegraphics[width=1\textwidth]{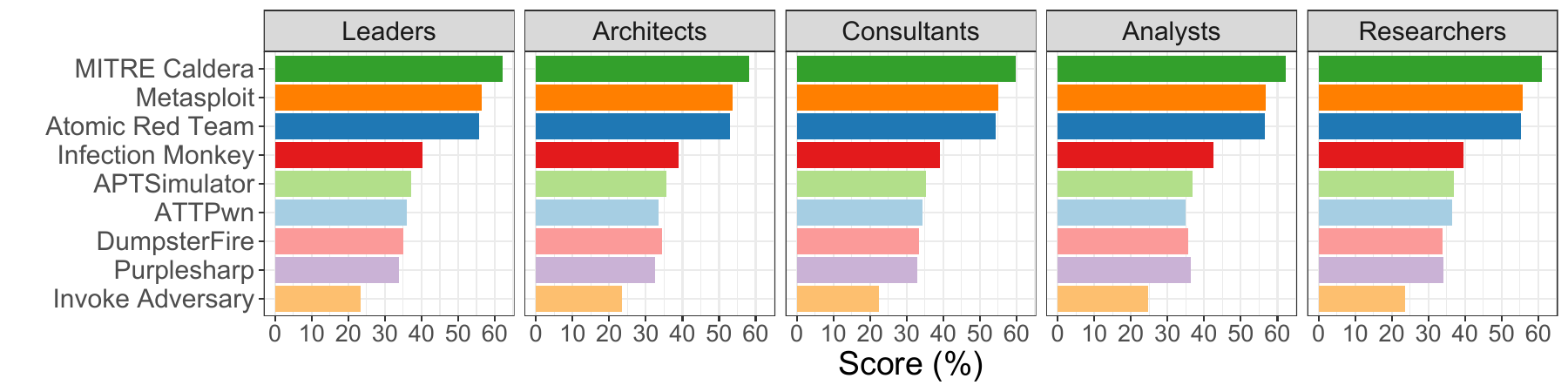}
		\caption{Scores of adversary emulation tools show that the ranks of tools are mostly the same across user roles.}
		\label{fig:scores}
		\vspace{-7pt}
	\end{figure*}
	
	Using Eq. (\ref{pw}) stated in Sect. \ref{ranking}, we are able to compute a single score for each tool and user role. Figure \ref{fig:scores} visualizes these scores and ranks the adversary emulation tools in ascending order according to their average score across all roles. As visible in the plot, MITRE Caldera is ahead of all tools but closely followed by Metasploit and Atomic Red Team that are roughly on par. These tools already emerged as top performers during our analysis of Sect. \ref{qualitative}, which suggests that weighting based on user requirements only has minor influence on the overall ranking. We also observe that Infection Monkey is slightly ahead, APTSimulator, ATTPwn, DumpsterFire, and Purplesharp all roughly achieve similar scores, and Invoke Adversary falls behind.
	
	\begin{figure*}
		\centering
		\includegraphics[width=1\textwidth]{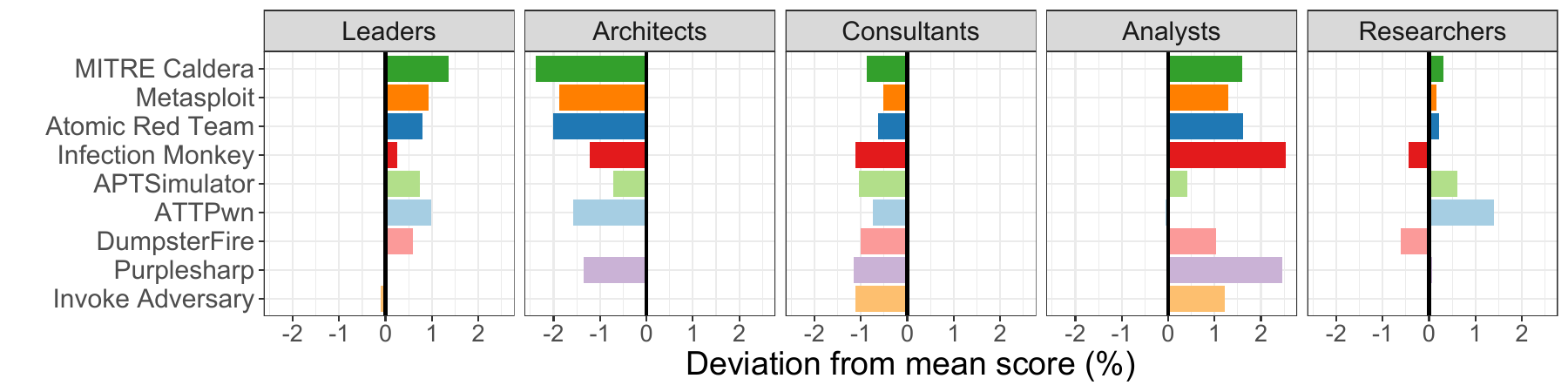}
		\caption{Deviations from the mean score shows which tools involve functions and properties that align with the requirements of specific user roles.}
		\label{fig:scoresdiff}
		\vspace{-7pt}
	\end{figure*} 
	
	The influence of weights from different user roles does not change the overall ranking of the top three tools. While the ranks of other tools change, this is mostly due to the fact that their average scores are close together and minor influences on the scores are sufficient to affect the ranking. Note that we mapped the responses of survey participants to quantitative weights in a static way (cf. Sect. \ref{ranking}), but this mapping can be changed to arbitrary magnitudes and thus have a stronger influence on the weighted scores. To obtain a better view on the influence of user roles on the tool ranking, we thus measure the deviation of each score from the mean score achieved across all roles and visualize the results in Fig. \ref{fig:scoresdiff}. The plots indicate that almost all tools are better suited to fulfill the needs of \textit{Leaders} and \textit{Analysts} than \textit{Architects} and \textit{Consultants}; there is no clear tendency for researchers. Moreover, functions and properties of some tools address the requirements from specific user groups, for example, Infection Monkey and Purplesharp yield significantly higher scores for \textit{Security Analysts} than for any other role, where scores are either below or around the average across all roles. We emphasize that this interpretation of the results only indicate that the tools are well suited for the requirements of specific user groups in comparison to an average user and not that they are necessarily the most suitable tools to be used by the respective user group. 
	
	\section{Discussion} \label{discussion}
	
	In this paper we outline a methodology to assess and rank adversary emulation tools based on technical evaluations and user requirements. Our questionnaire, survey results, and interpretations can be an assistance for organizations and security professionals when it comes to the selection of adequate tools for certain use-cases. Thereby, we recommend to consider updating the weights of relevant aspects so that they better reflect the requirements of the respective situations, for example, high usability (category \textit{U}) and availability of many predefined attack procedures (sub-category \textit{FC-7}) may be beneficial when the tool should be used by novices. 
	
	We recognize some limitations of our work. The selection of tools (cf. Sect. \ref{toolselection}) is based on public sources, which can be subject to bias. While we ensured to consult multiple sources, they may fail to include suitable tools that have only gained little attention or tools that have only been released very recently. Moreover, we notice that our selected tools vary strongly in terms of popularity (e.g., community size and number of forks on GitHub) and that tools with stronger public support usually outperform those that have fewer contributors or have even been abandoned by developers. In particular, comparing Fig. \ref{fig:stars} and Fig. \ref{fig:scores} shows that the ranking of tools we obtain from our evaluation mostly followed the ranking based on stars on GitHub. A notable exception to this observation is MITRE Caldera, which ranks first based on our evaluation but only fourth based on stars. Another limitation of our work is the small number of participants in our online survey. While we ensured that each group of users comprises at least 3 participants, we expect that a higher number of participants could yield more fine-granular insights into the differences between user roles. With these limitations in mind, we formulate the answers to our research questions as follows.
	
	\textit{RQ1: What properties of adversary emulation tools are the most relevant for stakeholders?} To answer this question, we worked out a questionnaire comprising categories and sub-categories of questions that address various aspects of adversary emulation tools. We then conducted a survey with domain experts to assess the relevance of each of the identified sub-categories. We present a detailed enumeration of user ratings in Sect. \ref{responses}, which shows that specific technical features, including automation, configuration, and execution of attack procedures, are among the most relevant features for stakeholders. We also found that the ability to create reports for results and absence of errors are important features of tools. 
	
	\textit{RQ2: Which adversary emulation tools are best suited to fulfill the needs of certain user groups?} Our analysis reveals that users have different priorities depending on their roles, for example, researchers prioritize technical capabilities over usability. We state the prioritized categories of questions for different user roles in Sect. \ref{responses} and analyze the influence of roles on the scores and ranking of adversary emulation tools in Sect. \ref{final}. Considering the weighted scores, we identify as MITRE Caldera, Metasploit, and Atomic Red Team as the most suitable tools across all user groups. These findings align with similar conclusions drawn from earlier studies \cite{zilberman2020sok}.
	
	\section{Conclusion} \label{conclusion}
	
	This paper presents a structured review and comparison of adversary emulation tools. Based on studies in related research areas, we design a questionnaire comprising five categories, namely \textit{Installation \& Configuration}, \textit{Community \& Support}, \textit{Documentation}, \textit{Usability}, and \textit{Features \& Capabilities}, with a total of 30 sub-categories and 80 questions. We select nine publicly available adversary emulation tools and evaluate to what degree they fulfill each of the questions. In addition, we conduct an online survey of domain experts to assign relevance scores to each sub-category of questions. Finally, we weight the evaluation results with user feedback to compute a single score for each tool. Our results suggest that independent from the user role, MITRE Caldera appears as the most suitable tool, followed by Metasploit and Atomic Red Team. For future work, we suggest to expand the range of adversary emulation tools to obtain more detailed insights into their peculiarities. In particular, we suggest to categorize tools and design questionnaires for each group to better capture and analyze properties that are specific to certain use-cases. Moreover, expert interviews that complement online surveys could be valuable to validate or come up with additional questions or categories of questions. Finally, while we only consider open-source tools in our survey, a similar study on commercial products could yield insights on specialized tools.
	
	\section*{Acknowledgment}
	
	Parts of this work were carried out in course of a Master's Thesis at the University of Applied Sciences Technikum Vienna \cite{mayer2024evaluierung}. The work in this paper has received funding from the European Union - European Defence Fund under GA no. 101103385 (AInception) and GA no. 101121403 (NEWSROOM). Views and opinions expressed are however those of the author(s) only and do not necessarily reflect those of the European Union. The European Union cannot be held responsible for them.
	
	
	\bibliographystyle{IEEEtran}
	\bibliography{bibliography}
	
	\appendix
	
	\subsection{Questionnaire} \label{appendix}
	
	This section contains the list of questions used to evaluate the selected adversary emulation tools in an offline setting. In the following, categories are written in bold, sub-categories are indicated by the number before the period, and question identifiers are indicated by the number after the period. We refer to Fig. \ref{fig:responses} for an enumeration of all sub-categories.
	
	\textbf{Installation \& Configuration (IC)}
	\begin{enumerate*}[label={\ding{226}}]
		\item IC-1.1: Which operating systems are supported? 
		\item IC-1.2: Are changes to security settings or permissions required?
		\item IC-2.1: Is third party software required?
		\item IC-2.2: Are there other requirements specific to operating systems?
		\item IC-3.1: Is it necessary to pre-configure the tool?
		\item IC-3.2: Are special network configurations required?
	\end{enumerate*}
	
	\textbf{Community \& Support (CS)}
	\begin{enumerate*}[label={\ding{226}}]
		\item CS-1.1: How many forks does the project have?
		\item CS-1.2: How many contributors does the project have?
		\item CS-1.3: What is the age of the project (in days)?
		\item CS-1.4: What is the size of the project (in kilobytes)?
		\item CS-2.1: How many programming languages does the project involve?
		\item CS-2.2: How many watchers does the project have?
	\end{enumerate*}
	
	\textbf{Documentation (D)}
	\begin{enumerate*}[label={\ding{226}}]
		\item D-1.1: Is there any kind of documentation for the tool?
		\item D-1.2: Is it sufficient for setting up all of the tool's components?
		\item D-1.3: Is it sufficient for launching built-in attacks?
		\item D-1.4: Is it sufficient for creating new custom attack procedures?
		\item D-1.5: Is it sufficient for creating new attack chains?
		\item D-1.6: Does the documentation describe how to interpret results of executed attack procedures?
		\item D-2.1: Is the available documentation updated with each release or other regular intervals?
		\item D-3.1: What is the average length of chapters?
		\item D-3.2: What is the average tree depth of chapters?
		\item D-4.1: What is the level of readability of the documentation?
		\item D-4.2: What is the medium length of sentences?
		\item D-4.3: How frequent are images or tables?
		\item D-5.1: If figures or tables are present, do they have an index number and caption?
		\item D-5.2: Is the documentation organized in accordance to the IEEE Standard for User Documentation?
	\end{enumerate*}
	
	\textbf{Usability (U)}
	\begin{enumerate*}[label={\ding{226}}]
		\item U-1.1: Are executions of processes consistent when tasks are performed multiple times?
		\item U-1.2: Is it possible to change tasks before the last step without starting the entire task from the beginning?
		\item U-1.3: Are there recurring errors that affect tasks and processes?
		\item U-2.1: Are interfaces and functions self-explanatory?
		\item U-2.2: Do users receive direct feedback for important intermediate steps and actions when using the tool?
		\item U-2.3: Is the appearance of the tool appealing? 
		\item U-3.1: Are there any interactive help functions other than static explanations of interfaces and functions?
		\item U-3.2: Are subsequent steps highlighted after completing a task?
		\item U-3.3: Are there help and information displays for buttons, functions, and interfaces?
		\item U-4.1: Does the tool support the creation of attack templates?
		\item U-5.1: Do all available user interfaces allow attack execution?
		\item U-5.2: Do all available user interfaces allow configuration of procedures?
		\item U-5.3: Do all available user interfaces allow to stop attack executions while they are still active?
		\item U-5.4: Do all available user interfaces allow to access log data of attack executions?
		\item U-5.5: Do all available user interfaces allow to create or add new custom procedures?
		\item U-5.6: Do all available user interfaces allow to create or add new attack chains?
		\item U-6.1: Is it possible to customize the user interface? 
		\item U-7.1: Is the average number of clicks or interactions appropriate for common tasks?
		\item U-7.2: Can tasks be solved by several approaches?
	\end{enumerate*}
	
	\textbf{Features \& Capabilities (FC)}
	\begin{enumerate*}[label={\ding{226}}]
		\item FC-1.1: Does the tool use agents? 
		\item FC-2.1: Are firewalls interrupting the workflow of the tool?
		\item FC-2.2: Are firewalls blocking the connection between the command-and-control server and the agent?
		\item FC-2.3: Is real-time antivirus interrupting remote access on the target system? 
		\item FC-2.4: Is real-time antivirus interrupting the workflow of the tool?
		\item FC-2.5: Is real-time antivirus interrupting the functionality of the agent?
		\item FC-2.6: Is real-time antivirus deleting the script or interrupting its functionality?
		\item FC-2.7: Is real-time antivirus interrupting functionality of third party tools?
		\item FC-3.1: Does the tool support cleanup of the target system after attack?
		\item FC-3.2: Does cleanup occur immediately after the relevant attack procedure?
		\item FC-3.3: Is cleanup of the relevant procedures only possible at the end of an attack chain?
		\item FC-4.1: Does the tool have logging capabilities?
		\item FC-4.2: Is every executed procedure logged during an attack?
		\item FC-4.3: How is the result of attacks presented? 
		\item FC-5.1: Does the tool support repeated execution of the same attack procedure but with different parameters?
		\item FC-5.2: Are custom attack procedures created and handled in the same way as predefined procedures?
		\item FC-5.3: Is it possible to resume incomplete attack procedures or actions after restarting the app?
		\item FC-6.1: Is parallel execution of attack procedures supported?
		\item FC-6.2: It is possible to automatically execute several attack procedures one after another?
		\item FC-7.1: Are the instructions provided within the tool sufficient so that even a novice is able to execute built-in attack procedures?
		\item FC-7.2: Are predefined attack chains that consist of several attack procedures available in the tool?
		\item FC-7.3: What range of tactics and techniques from MITRE ATT\&CK are covered by predefined attack procedures?
		\item FC-7.4: Is the tool designed to support blue teaming?
		\item FC-8.1: Is it possible to reconfigure predefined attack procedures?
		\item FC-8.2: Is it possible to create new custom attack chains?
		\item FC-8.3: Is it possible to create new custom attack procedures through any of the interfaces?
		\item FC-8.4: Is it possible to add new custom attack procedures as scripts?
		\item FC-9.1: Is it possible to stop and reconfigure an ongoing attack at any point in time?
		\item FC-10.1: Are special privileges required for the agents deployed on the target systems?
		\item FC-10.2: Are special privileges required for any of the scripts?
		\item FC-10.3: Are special privileges required for any of the involved third party tools?
		\item FC-11.1: Does the tool execute scripts on the endpoints? 
		\item FC-11.2: Does the tool support attack scripting?
		\item FC-11.3: Are attack procedures implemented using scripts?
	\end{enumerate*}
	
\end{document}